\documentclass[prc]{revtex4}

\usepackage{graphicx,amsmath}
\begin{document}
\title{Formulating Viscous Hydrodynamics for Large Velocity Gradients}
\author{Scott Pratt}
\affiliation{Department of Physics and Astronomy,
Michigan State University\\
East Lansing, Michigan 48824-1321}
\date{\today}

\begin{abstract}
Viscous corrections to relativistic hydrodynamics, which are usually formulated for small velocity gradients, have recently been extended from Navier-Stokes formulations to a class of treatments based on Israel-Stewart equations. Israel-Stewart treatments, which treat the spatial components of the stress-energy tensor $\tau_{ij}$ as dynamical objects, introduce new parameters, such as the relaxation times describing non-equilibrium behavior of the elements $\tau_{ij}$. By considering linear response theory and entropy constraints, we show how the additional parameters are related to fluctuations of $\tau_{ij}$. Furthermore, the Israel-Stewart parameters are analyzed for their ability to provide stable and physical solutions for sound waves. Finally, it is shown how these parameters, which are naturally described by correlation functions in real time, might be constrained by lattice calculations, which are based on path-integral formulations in imaginary time.
\end{abstract}


\maketitle

\section{Introduction and Basic Theory}

The liquid-like quark matter observed at the Relativistic Heavy Ion Collider (RHIC) \cite{rhicwhitepapers} has inspired a renaissance in hydrodynamic modeling at high energy. However, even though the matter is in a liquid state, the extremely short characteristic time and distance scales necessitate the inclusion of viscous effects in hydrodynamic models. Viscous effects, or equivalently the effects of non-equilibrium values of the spatial components of the stress-energy tensor $\tau_{ij}$, are especially important during the first few fm/$c$ of the collision which is characterized by large velocity gradients and enormous shears. Bulk viscosity might play an important role during the hadronization stage, arising from the matter's inability to maintain equilibrium when traversing the phase transition \cite{pratt_sqm,paech,kharzeev,romatshcke,songheinz}. 

The standard means for implementing viscous effects into hydrodynamics is the Navier-Stokes (NS) equations \cite{weinberg}. However, during the last few years, and especially for relativistic heavy ion physics, increasing attention has been focused on the Israel-Stewart (IS) equations \cite{israelstewart,muronga,baierromatschke,heinzchaudhuri}. Whereas the spatial components of the stress-energy tensor are determined by the energy density and velocity gradients in NS theory, in IS treatments the components are treated as dynamic objects which decay exponentially towards the NS values. In this paper the IS equations are analyzed in detail, the differences between IS and NS descriptions are explored, and the degree to which the additional parameters required for IS theory might be microscopically calculable from approaches like lattice gauge theory is investigated. After reviewing the basic motivations and premises of IS-based equations in this section, the formulations are analyzed from the perspectives of entropy growth and linear response theory in Sec. \ref{sec:entropy}. These perspectives constrain both the form and parameters of IS theory. In particular, the relaxation times, along with the viscosities, are shown to determine both the functional form for the relaxation and the statistical fluctuation of the elements of the stress-energy tensor. Some of these constraints are also derived from the perspective of linear response theory and are presented in the subsequent section. Section \ref{sec:largedeviations} presents a discussion of how one might alter IS equations to be more physically relevant for large deviations from equilibrium, and Sec. \ref{sec:sonic} provides an analysis of the range of IS parameters that provide stable and physical sound waves. The prospects for determining IS parameters from lattice calculations are discussed in \ref{sec:pathintegrals} .

The traditional means for incorporating viscosity is through the Navier-Stokes equation, in which the stress-energy tensor is modified by the velocity gradients in a linear fashion,
\begin{eqnarray}
\tau_{ij}=P\delta_{ij}-\eta\left[\partial_iv_j+\partial_jv_i-(2/3)\nabla\cdot {\bf v}\right]
-\zeta\delta_{ij}\nabla\cdot v.
\end{eqnarray}
Here, it has been assumed that the stress-energy tensor $\tau_{ij}$ is expressed in the rest-frame of the fluid, where ${\bf v}=0$. The pressure $P$, the shear viscosity $\eta$, and the bulk viscosity $\zeta$ are all functions of the energy and particle densities, though we will ignore the particle-density dependence throughout the remainder of the paper. Since the stress-energy tensor is symmetric, the deviation of $\tau_{ij}$ from the equilibrium value $P\delta_{ij}$ can be expressed in terms of six independent numbers:
\begin{eqnarray}
\label{eq:abdef}
b&\equiv&\frac{1}{3}\left(\tau_{xx}+\tau_{yy}+\tau_{zz}\right)-P,\\
\nonumber
a_1&\equiv&\frac{1}{2}\left(\tau_{xx}-\tau_{yy}\right),\\
\nonumber
a_2&\equiv&\frac{1}{\sqrt{12}}\left(\tau_{xx}+\tau_{yy}-2\tau_{zz}\right),\\
\nonumber
a_3&\equiv&\tau_{xy},~~a_4\equiv\tau_{xz},~~a_5\equiv\tau_{yz}.
\end{eqnarray}
Here, $b$ refers to the deviation from equilibrium of the trace of $\tau_{ij}$ and is related to the bulk viscosity, while the five values $a_i$, which are related to the shear viscosity, describe the angular anisotropy. In addition to $\nabla\cdot v$ one can also define five symmetrized ``velocity gradients'',
\begin{eqnarray}
\label{eq:wdef}
\omega_1&\equiv&\frac{1}{2}\left(2\partial_xv_x-2\partial_yv_y\right),\\
\nonumber
\omega_2&\equiv&\frac{1}{\sqrt{3}}\left(\partial_xv_x+\partial_yv_y-2\partial_zv_z\right),\\
\nonumber
\omega_3&\equiv&\left(\partial_xv_y+\partial_yv_x\right),~~ \omega_4\equiv\left(\partial_xv_z+\partial_zv_x\right),~~ \omega_5\equiv\left(\partial_yv_z+\partial_zv_y\right).
\end{eqnarray}
With these definitions, the work (in the local rest frame) due to expanding a volume $V$ for a time increment $dt$ becomes,
\begin{equation}
\label{eq:dw}
\frac{1}{V}dW=dt\sum_{ij}\tau_{ij}\partial_iv_j=dt\left[P(\nabla\cdot v)+b\nabla\cdot v +\sum_i a_i\omega_i\right],
\end{equation}
and the Navier-Stokes equations assume a simple form,
\begin{eqnarray}
\label{eq:ns}
b^{({\rm NS})}&=&-\zeta\nabla\cdot v,\\
\nonumber
a_i^{({\rm NS})}&=&-\eta \omega_i.
\end{eqnarray}

Israel-Stewart approaches \cite{israelstewart} differ from NS in that $a_i$ and $b$ are treated as dynamical objects. Originally, the motivation for IS approaches was related to their numerical stability compared to NS equations \cite{muronga}. This property follows from the fact that the equations are hyperbolic, as opposed to NS equations which are parabolic, similar to the diffusion equation. Since the diffusion current can move arbitrarily quickly for arbitrarily large gradients, the equations were effectively non-causal. Though the violation of causality was never shown to be important in the context of RHIC collisions, IS solutions have been considered safer than NS solutions. An analysis of the stability of sonic modes of a given wavelength is presented in Section \ref{sec:sonic}.

The basic premise of IS solutions is that the offsets of $\tau_{ij}$ decay toward their NS values exponentially,
\begin{eqnarray}
\label{eq:isassumption}
\frac{D}{Dt}\left(\frac{a_i}{\alpha}\right)=-\frac{1}{\alpha}\left({a_i}-a_i^{({\rm NS})}\right)/\tau_a,\\
\nonumber
\frac{D}{Dt}\left(\frac{b}{\beta}\right)=-\frac{1}{\beta}\left(b-b^{({\rm NS})}\right)/\tau_b,
\end{eqnarray}
where $\alpha$ and $\beta$ are functions of the energy density. They are usually set to unity, and if the energy density is fixed, the forms for $\alpha$ and $\beta$ are irrelevant.  Providing an energy dependence to $\alpha$ and $\beta$ change the equations of motion for large velocity gradients. For instance, with a large velocity gradient a system will expand by a given fraction in a time inversely proportional to the velocity gradients. If the expansion time is much less than the relaxation times $\tau_a$ and $\tau_b$, the ratios $a_i/\alpha$ and $b/\beta$ will be effectively frozen for very rapid expansions. As an example, if $\alpha$ were set equal to $P$, a very rapid expansion would freeze the ratios $a_i/P$, whereas setting $\alpha=1$ would lead to a freezing out of the variation from equilibrium $a_i$. For a rapidly cooling system where $P$ is rapidly falling, the choice of $\alpha=P$ might keep the components of $\tau_{ij}$ positive, whereas the choice $\alpha=1$ might lead to negative elements.

As an aside, we mention that the derivatives $D/Dt$ in Eq.s (\ref{eq:isassumption}) include the effects of rotational flow \cite{baierromatschke}. For a tensor $\tau_{ij}$,
\begin{eqnarray}
\label{eq:rotation}
\frac{D}{Dt}\tau_{ij}&=&\frac{\partial}{\partial t}\tau_{ij}+\frac{1}{2}\left[\Omega,\tau\right]_{ij},\\
\nonumber
\Omega_{ij}&\equiv&\frac{\partial_iv_j-\partial_j v_i}{2}~.
\end{eqnarray}
This differs from the usual notation where $D/Dt$ also includes the term ${\bf v}\cdot\nabla$. However, that term is explicitly zero in the rest frame of the fluid. If one were solving the equations of motion for $\Omega\ne 0$, one would need to rotate the stress energy tensor at every time step consistently with Eq. (\ref{eq:rotation}). This would mix the five elements $a_i$, but would leave $b$ unchanged. Since $\Omega$ is proportional to velocity gradients, and since $a_i$ is also proportional to viscosity gradients in NS theory, corrections for rotational flow are considered a ``second-order'' correction from the NS perspective. Thus, one expects rotational effects to be negligible except for those instances where there exists both rotation and significant values for the anisotropies $a_i$. Henceforth, we will neglect rotation, though we emphasize that it might become non-negligible for non-central collisions, or for flow away from mid-rapidity.

The parameters of IS hydrodynamics are the viscosities, $\eta$ and $B$, the relaxation times, $\tau_a$ and $\tau_b$, and the scaling functions for the exponential decays, $\alpha$ and $\beta$. Each of these six parameters can be functions of the energy density. However, as will be seen in the next section, the constraint that the entropy cannot fall constrains the number of parameters to four, and relates them to the statistical fluctuations of $\tau_{ij}$.

\section{Entropy Constraints}
\label{sec:entropy}

Entropy must rise, and since the IS equations describe entropy production, the IS parameters are constrained. Furthermore, the entropy should depend not only on the energy density, but also on the deviation of $\tau_{ij}$ from equilibrium. In this section we will show how the entropy constraint reduces the number of IS parameter from six to four, and relates them to the statistical fluctuation of $\tau_{ij}$, which in turn is related to how the entropy is affected by non-zero values of $a_i$ and $b$.

For a fixed energy density, $\tau_{ij}$ should equal $P$ if the entropy is maximized. For small variations, $a_i$ and $b$, the entropy penalty will grow quadratically with $a$ and $b$.
\begin{equation}
s=s_{\rm equil}(\epsilon)-\frac{b^2}{2\sigma_b^2}-\frac{\sum_{i}a_{i}^2}{2\sigma_a^2},
\end{equation}
where $\sigma_a$ and $\sigma_b$ are functions of the energy density. Since the probability for a given fluctuation of the stress energy tensor occurs with probability $e^S=e^{sV}$, the variances for $a_i$ and $b$ in a volume $V$ are
\begin{eqnarray}
\langle b^2 \rangle= \frac{\sigma_b^2}{V},~~~~ \langle a_i^2\rangle=\frac{\sigma_a^2}{V}.
\end{eqnarray}
The fact that the fluctuations of $a_i$ and $b$ fall as $1/\sqrt{V}$ is expected since the correlations  persist over a finite domain and the number of independent domains should increase linearly with $V$. Equivalently, one would expect the fluctuation of the quantities $a_iV$ and $bV$ to increase with $\sqrt{V}$.

The change in entropy is then
\begin{equation}
dS = dS_{\rm equil}-Vdt \left[
\frac{b}{\sigma_b}\frac{d}{dt}\left(\frac{b}{\sigma_b}\right)
+\frac{a_i}{\sigma_a}\frac{d}{dt}\left(\frac{a_i}{\sigma_a}\right)
\right].
\end{equation}
The change in the equilibrium entropy for an expanding volume element is
\begin{equation}
dS_{\rm equil}=(dE+PdV)/T.
\end{equation}
where $P$ and $T$ are functions of $\epsilon$ assuming equilibrium.

After using the fact that $dV=Vdt(\nabla\cdot v)$, and using Eq. (\ref{eq:dw}) for the differential work, which equals $dE$,
\begin{equation}
\label{eq:srate}
\frac{T}{V}\frac{dS}{dt}= -b(\nabla\cdot v) -\sum_i a_i\omega_i
-T \frac{b}{\sigma_b}\frac{d}{dt}\left(\frac{b}{\sigma_b}\right)
-T \sum_i \frac{a_i}{\sigma_a}\frac{d}{dt}\left(\frac{a_i}{\sigma_a}\right).
\end{equation}
As expected, entropy production vanishes for $a_i=b=0$.

Inserting the IS equations of motion in Eq. (\ref{eq:isassumption}) into Eq. (\ref{eq:srate}),
\begin{eqnarray}
\label{eq:dsdt}
\frac{T}{V}\frac{dS}{dt}&=&b\nabla\cdot v\left[\frac{T\zeta}{\sigma_b^2\tau_b}-1\right]
+\sum_i a_i\omega_i\left[\frac{T\eta}{\sigma_b^2\tau_a}-1\right]\\
\nonumber
&&+\frac{b^2T}{\sigma_b^2}\left[\frac{1}{\tau_b}-\frac{d}{dt}\ln(\beta/\sigma_b)\right]
+\sum_i\frac{a_i^2T}{\sigma_a^2}\left[\frac{1}{\tau_a}-\frac{d}{dt}\ln(\alpha/\sigma_a)\right].
\end{eqnarray}

For the entropy to rise regardless of the values of $a$ and $b$ or the velocity gradients, the terms linearly proportional to $a_i$ and $b$ must disappear as well as those linearly proportional to the velocity gradients $\omega_i$ and $\nabla\cdot v$. Furthermore, since $\alpha$ and $\beta$ are functions of the energy and can change arbitarily quickly for arbitrarily rapid expansions, ensuring that the entropy always rises requires stating that both $\alpha/\sigma_a$ and $\beta/\sigma_b$ are constants, which can be set to unity without changing any behavior,
\begin{equation}
\alpha=\sigma_a,~~~~\beta=\sigma_b.
\end{equation}
If the terms linear in $a$ and $b$ are to never contribute to the entropy regardless of the values of $a$ and $b$,
\begin{equation}
\label{eq:alphabetaresult}
\sigma_a^2=\frac{T\eta}{\tau_a}, ~~~\sigma_b^2=\frac{T\zeta}{\tau_b}.
\end{equation}
Thus, the viscosities and relaxation times uniquely determine the scaling functions for the exponential decays, $\alpha$ and $\beta$, which are equivalent to the fluctuations of $\sigma_a$ and $\sigma_b$ at fixed energy. This latter expression was derived from the perspective of the Boltzmann equation \cite{muronga}. This derivation is both more general, and provides the equivalence between the fluctuations and the scaling functions.

After the parameter constraints are enforced, the growth rate for the entropy has a simple form,
\begin{equation}
\label{eq:sproduction}
\frac{T}{V}\frac{dS}{dt}=\frac{b^2}{\zeta}+\sum_i\frac{a_i^2}{\eta}.
\end{equation}
After enforcing the NS equations in Eq. (\ref{eq:ns}), this reproduces the usual textbook result that the rate of entropy production is proportional to $(\nabla\cdot{\bf v})^2$ and $\omega_i^2$. The crucial difference between the entropy production rate in Eq. (\ref{eq:sproduction}) and the equivalent NS result becomes important in the limit of a very rapid expansion. The time required for the dimensions of the volume to increase by a specific amount will be proportional to $1/(\nabla\cdot{\bf v})$ or $1/\omega_i$. Considering a time step,
\begin{equation}
\Delta t = \gamma/\nabla\cdot v,
\end{equation}
a fluid element will increase by a factor,
\begin{equation}
V(t+\Delta t)=V(t)e^\gamma.
\end{equation}
For a small $\gamma$, the equations of motion for $a_i/\sigma_a$ and $b/\sigma_b$ in Eq. (\ref{eq:isassumption}) show that in the limit of large velocity gradients they change by an amount:
\begin{equation}
\Delta(b/\sigma_b)=\gamma \zeta/\tau_b,~~~\Delta a_i/\sigma_a=\gamma\eta\frac{\omega_i}{(\nabla\cdot v)\tau_a},
\end{equation}
which, as Eq. (\ref{eq:sproduction}) shows, results in the entropy production proportional to $\gamma$. Thus, in the limit of infinitely fast velocity gradients, the net entropy production approaches a fixed value depending on $\gamma$. In contrast, the NS production rate is proportional to the velocity gradients squared, so that the net entropy produced for a fixed expansion factor $\gamma$ is divergent as $\nabla\cdot v\rightarrow\infty$. For the opposite extreme, of a small velocity gradient, the values of $b$ and $a_i$ in Eq. (\ref{eq:sproduction}) will relax to the NS values, and both the IS and the NS results for entropy production will be identical.

\section{From the Perspective of Linear Response Theory}
\label{sec:linearresponse}

According to linear response theory \cite{forster,paech}, the classical limit of the Kubo formula gives,
\begin{eqnarray}
\eta&=&\frac{1}{T}\int_0^\infty dr_0 \int d^3r \left\langle \tau_{ij}(0)\tau_{ij}(r)\right\rangle,
~~~i\ne j\\
\nonumber
&=&\frac{1}{T}\int_0^\infty dr_0\int d^3r \left\langle a_i(0)a_i(r)\right\rangle, ~~~{\rm any}~i,\\
\nonumber
\zeta&=&\frac{1}{9T}\sum_{i,j}\int_0^\infty dr_0 \int d^3r \left\langle \left(\tau_{ii}(0)\tau_{jj}(r)-P^2\right)\right\rangle,~~~i\ne j\\
\nonumber
&=&\frac{1}{T}\int_0^\infty dr_0\int d^3r \left\langle b(0)b(r)\right\rangle.
\end{eqnarray}
Here, the averages involve summing over states at fixed energy. If the energy is allowed to vary, as in a grand canonical ensemble, one must subtract the contribution from fluctuating energy, $c_s^4\langle \delta E^2\rangle$, from the expression for the bulk viscosity. 

For a fluctuating field that loses correlation exponentially, $\sim e^{-t/\tau_a}$, 
\begin{eqnarray}
\eta&=&\frac{\tau_a V}{T} \left\langle \bar{a}_i^2\right\rangle,~~~\bar{a}_{i}\equiv \frac{1}{V}\int d^3r ~a_i(r),\\
\nonumber
\zeta&=&\frac{\tau_b  V}{T} \left\langle \bar{b}^2\right\rangle,~~~\bar{b}\equiv \frac{1}{V}\int d^3r ~b(r).
\end{eqnarray}
Given that the entropy penalty for small $a^2$ is,
\begin{equation}
\label{eq:ehatS}
e^S\approx e^{S_{\rm equil}}\exp\left\{-\int d^3r~ \left[\sum_i(\bar{a}_i^2/2\sigma_a^2)
+\int d^3r~(\bar{b}^2/2\sigma_b^2)\right]\right\},
\end{equation}
one can identify the variances of $\bar{a}$ and $\bar{b}$,
\begin{eqnarray}
\left\langle\bar{a}_i^2\right\rangle&=&\frac{\sigma_a^2}{V},\\
\nonumber
\left\langle\bar{b}^2\right\rangle&=&\frac{\sigma_b^2}{V}.
\end{eqnarray}
The $1/V$ factor was expected given that there are no long-range correlations in $\tau_{ij}$.

Plugging this into the expression for the viscosities,
\begin{equation}
\label{eq:flucdis}
\eta=\frac{\tau_a\sigma_a^2}{T},~~~~\zeta=\frac{\tau_b\sigma_b^2}{T}
\end{equation}
which matches the result of the last section.

\section{Large deviations from Equilibrium}
\label{sec:largedeviations}

As shown in the previous sections, IS formalism handles arbitrarily large velocity gradients, but remains an expansion in the deviations $a_i$ and $b$. The non-equilibrium values $a/\sigma_a$ and $b/\sigma_b$ in Eq. (\ref{eq:isassumption}) decay towards values which can be arbitrarily large for arbitrarily large velocity gradients. However, this become unphysical in some cases as the elements $\tau_{ij}$ become arbitrarily large, or strongly negative. Depending on the source of the viscosity, it might be more physical to impose a limit on the deviation of $\tau_{ij}$ from equilibrium. For instance, if the shear viscosity is caused by a finite mean free path, the kinetic pressure should remain positive, whereas if the dynamics are those of classical electric fields, the stress-energy tensor should be confined to the region, $-\epsilon< \tau_{ij}< \epsilon$. Here, we provide a simple recipe to enforce such constraints, within the IS picture, that do not violate entropy constraints. 

First we consider the case of the bulk viscosity, and assume that there exist some physical constraints that enforce $|b|<b_{\rm max}$. To dynamically limit $b$, we alter the equations of motion,
\begin{eqnarray}
\frac{dx}{dt}&=&-\left[x- (\zeta\nabla\cdot v)/\sigma_b\right]/\tau_b,\\
\nonumber
b&=&b_{\rm max}\tanh\left(\frac{\sigma_b x}{b_{\rm max}}\right).
\end{eqnarray}
For small $b$ these become identical to the previous IS equations of motion. The equations for entropy production become,
\begin{equation}
\frac{T}{V}\frac{dS}{dt}=b\nabla\cdot v-T\frac{ds}{dx}\left[x- (\zeta\nabla\cdot v)/\sigma_b\right]/\tau_b.
\end{equation}
Again, the term proportional to $\nabla\cdot v$ must vanish if the entropy is to always grow despite the sign of the velocity gradient or the value of $x$. This determines $ds/dx$,
\begin{equation}
\frac{ds}{dx}=-b\frac{\sigma_b\zeta}{T\tau_b}=-\frac{b}{\sigma_b},
\end{equation}
where Eq. (\ref{eq:alphabetaresult}) was used to simplify the prefactor. Integrating to find the entropy density,
\begin{equation}
s=s_{\rm eq}-\frac{b_{\rm max}^2}{\sigma_b^2}\ln\left[\cosh(\sigma_b x/b_{\rm max})\right].
\end{equation}
The rate of entropy production is then manifestly positive,
\begin{equation}
\frac{T}{V}\frac{dS}{dt}=T\frac{b}{\sigma_b\tau_b}x=xT\frac{b_{\rm max}}{\sigma_b\tau_b}\tanh(\sigma_b x/b_{\rm max}),
\end{equation}
which by virtue of Eq. (\ref{eq:flucdis}) becomes
\begin{equation}
\frac{T}{V}\frac{dS}{dt}=\frac{bx\sigma_b}{\zeta}=\frac{x\sigma_b}{\zeta}\tanh(\sigma_bx/b_{\rm max})
=\frac{bb_{\rm max}}{\zeta}\tanh^{-1}(b/b_{\rm max})~.
\end{equation}
Again, for small $b$ this gives the identical behavior as the previous sections.

A similar procedure can be followed for the shear terms to enforce the constraint,
\begin{equation}
a_1^2+a_2^2<a_{\rm max}^2,
\end{equation}
by assuming equations of motion,
\begin{eqnarray}
\frac{dy_i}{dt}&=&-\frac{1}{\tau_a}\left(y_i-\eta\omega_i/\sigma_a\right),\\
\nonumber
a&=&a_{\rm max}\tanh\left(\frac{\sigma_ay}{a_{\rm max}}\right),~~y=\sqrt{y_1^2+y_2^2},\\
\nonumber
a_i&=&a\frac{y_i}{y},~~~a=\sqrt{a_1^2+a_2^2}.
\end{eqnarray}
After enforcing the constraint that the entropy grows regardless of the value of $\omega_i$, one comes to the result for both shear and bulk deviations from equilibrium,
\begin{eqnarray}
\frac{T}{V}\frac{dS}{dt}&=&\frac{x\sigma_b}{\zeta}\tanh(x\sigma_b/b_{\rm max})
+\frac{y\sigma_a}{\sigma_a}\tanh(y\sigma_a/a_{\rm max}),\\
\nonumber
s&=&s_{\rm eq}-\frac{b_{\rm max}^2}{\sigma_b^2}\ln\left[\cosh(\sigma_b x/b_{\rm max})\right]
-\frac{a_{\rm max}^2}{\sigma_a^2}\ln\left[\cosh(\sigma_a y/a_{\rm max})\right].
\end{eqnarray}
The expression for entropy generation is equivalent to that described in Eq. (\ref{eq:sproduction}) in the limit of small $a_i/a_{\rm max}$ and $ b/b_{\rm max}$.

This procedure is more phenomenology than theory. Whereas the coefficients $\eta$, $\zeta$, $\sigma_a$ and $\sigma_b$ can all be formally expressed in terms of microscopic correlations, the additional parameters $a_{\rm max}$ and $b_{\rm max}$, as well as the formalism itself, are chosen at the discretion of the modeler. In fact, one might determine the parameters, or alter the formalism, based on a numerical comparison of the IS evolution described above with a particular microscopic model. For instance, one might run a model based on particular color-glass assumptions in a simplified geometry, then tune the IS approach to match the behavior of the stress-energy tensor. The IS model could then be considered a surrogate for the color-glass model, but would more naturally couple to the more theoretically justified hydrodynamic treatment of the later stage, and also more naturally incorporate complicated three-dimensional evolution.

\section{Stability of Israel-Stewart Solutions}
\label{sec:sonic}

One of the often-stated advantages of IS equations derives from the numerical stability of the approach \cite{muronga}. Furthermore, the solutions to NS equations can contain super-luminar transport. Here, solutions to IS equations are presented for small perturbations. After solving for the angular frequency, $\omega(k)$, the group velocity $d\omega/dk$ is plotted. In the limit of the relaxation times becoming zero, the IS and NS solutions become identical, and high momentum modes become super-luminar or unstable. For longer relaxation times, the solutions are well behaved, and a characteristic scale for the transition from poorly- to well-behaved can be expressed in terms of the viscosity, speed of sound, and enthalpy.

To solve for the sound waves, we first consider perturbations of the energy density $\delta\epsilon$, the velocity $v$ and the deviation of the stress-energy tensor, $\delta \tau_{zz}$.
\begin{eqnarray}
\label{eq:planewave}
\delta\epsilon&=&Ee^{i\omega t+ikz},\\
\nonumber
\delta \tau_{zz}&=&Ae^{i\omega t+ikz},\\
\nonumber
v_z&=&Ve^{i\omega t+ikz}
\end{eqnarray}
The IS equations of motion for small $E$, $A$ and $V$ are:
\begin{eqnarray}
h\frac{\partial v_z}{\partial t}&=&-\frac{d}{dz}\left(c_s^2\delta\epsilon+\delta \tau_{zz}\right),\\
\nonumber
\frac{\partial \tau_{zz}}{\partial t}&=&-\frac{1}{\tau_b}\left(\tau_{zz}+\zeta\frac{\partial v_z}{\partial z}\right),
\\
\nonumber
\frac{\partial\delta\epsilon}{\delta t}&=&-h\frac{\partial v}{\partial z}.
\end{eqnarray}
Using the plane-wave form of Eq. (\ref{eq:planewave}), and assuming $\eta=0$, the IS equations of motion become:
\begin{eqnarray}
\label{eq:ISsound}
\omega E&=&-hkV,\\
\nonumber
\omega A&=&\frac{i}{\tau_b}A-\frac{k\zeta}{\tau_b}V,\\
\nonumber
\omega V&=&-\frac{kc_s^2}{h}E-\frac{k}{h}A.
\end{eqnarray}
One can then solve these for $x\equiv i\omega/kc$ and obtain a cubic equation,
\begin{eqnarray}
\label{eq:cubic}
x^3+Cx^2+(1+\gamma)x+C&=&0,\\
\nonumber
C&\equiv&\frac{1}{kc_s\tau_b},\\
\nonumber
\gamma&\equiv&\frac{\zeta}{hc_s^2\tau_b}.
\end{eqnarray}
These approach the NS equations when $\tau_b\rightarrow 0$.

\begin{figure}
\centerline{\includegraphics[width=2.8in]{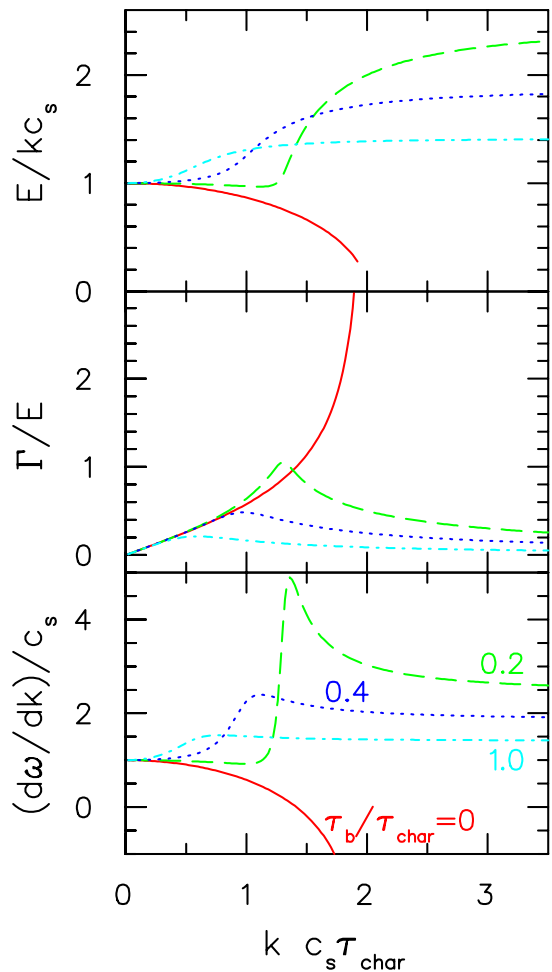}}
\caption{\label{fig:sound}(color online)
As a function of the wave number $k$, the real and imaginary solutions for $\omega(k)=E+i\Gamma$ are shown in the upper two panels, while the phase velocity, $d\omega/dk$, is shown in the lower panel. By scaling the wave number by the characteristic time, defined in Eq. (\ref{eq:chartimedef}), the solutions depend only on the ratio of the relaxation time $\tau_b/\tau_{\rm char}$. For $\tau_b\rightarrow 0$ (solid line), the solutions approach Navier-Stokes solutions, which give peculiar behavior for wave numbers larger than $1/c_s\tau_{\rm char}$. The behavior ameliorates for higher relaxation times, $\tau_{\rm b}/\tau_{\rm char}=0.2$ (dashes), $\tau_{b}/\tau_{\rm char}=0.4$ (dotted), $\tau_b/\tau_{\rm char}=1.0$ (dot-dashed).}
\end{figure}

The solutions to Eq.s (\ref{eq:cubic}) are shown in Fig. \ref{fig:sound}. The three panels illustrate the solutions as a function of the wave number $k$ for the real and imaginary parts of $\omega$, and the group velocity $d\omega/dk$. To provide a more general expression, frequencies are given in units of the inverse characteristic time,
\begin{equation}
\label{eq:chartimedef}
\tau_{\rm char}\equiv \frac{\zeta}{hc_s^2},
\end{equation}
which has dimensions of time provided $c_s$ is given as a fraction of the speed of light. This time characterizes the strength of the viscous term. The three curves then depend only on the ratio of the relaxation times to the characteristic times, $\tau_b/\tau_{\rm char}$. For small relaxation times, $\omega(k)$ becomes negative and diverges, leading to arbitrarily large negative phase velocities. For $\tau_\zeta/\tau_{\rm char}>\sim 0.16$, the divergence disappears, but solutions remain rather peculiar in that the speed of sound remains super-luminar for certain wavelengths. For larger relaxation times the behavior seems normal, and the phase velocities only modestly differ from the nominal speed of sound. Not unexpectedly, in the same region where the phase velocities most greatly differ from the speed of sound, the imaginary part of $\omega$, which describes the decay of the sound wave, becomes large. 

Since the solutions $\omega(k)$ were generated by solving a cubic equation, there are three solutions. Two of them correspond to plane waves moving in the positive and negative directions, and are described by the solutions plotted in Fig. \ref{fig:sound}. For the third solution, the real part of $\omega$ is zero, and the waves do not propagate, and only decay.

The solutions presented here are for the case with only bulk viscosity. One can solve the analogous set of equations as (\ref{eq:ISsound}) for having shear viscosity, but no bulk viscosity. In that case, Eq.s (\ref{eq:abdef}) give $\delta \tau_{zz}=(2/\sqrt{3})a_2$, and given the Eq.s (\ref{eq:wdef}), the IS equations of motion for $a_2$ are
\begin{equation}
\frac{\partial a_2}{dt}=-\frac{1}{\tau_a}\left[a_2+(2/\sqrt{3})\eta\partial_zv_z\right].
\end{equation}
Finding $\omega(k)$ for the sonic waves gives the same result as shown in Fig. \ref{fig:sound}, only with the definition,
\begin{equation}
\tau_{\rm char}=\frac{4\eta}{3hc_s^2}.
\end{equation}
If one were to simultaneously consider both shear and bulk terms, one would have to simultaneously solve four equations of motion, rather than the three given in Eq. (\ref{eq:ISsound}). This would lead to a quartic equation, unless the two relaxation times were equal.

\section{Determining Coefficients from Path Integrals}
\label{sec:pathintegrals}

The transport coefficients used in IS theory can all be expressed in the terms of correlations. Using the full quantum expressions \cite{paech},
\begin{eqnarray}
\label{eq:transtheory}
\sigma_a^2&=&\int d^3r~\left\langle T_{xy}(0)T_{xy}(t=0,{\bf r})\right\rangle,\\
\nonumber
\eta&=&\frac{-i}{2}\int d^4x~x_0 \left\langle \left[T_{xy}(0),T_{xy}(x)\right]\right\rangle,\\
\nonumber
\sigma_b^2&=&\frac{1}{3}\int d^3r~
\left\langle\delta\left(T_{xx}(0)+T_{yy}(0)+T_{zz}(0)\right)\delta T_{xx}(t=0,{\bf r})\right\rangle,\\
\nonumber
\zeta&=&\frac{-i}{6} \int d^4x~
x_0 \left\langle \left[\delta\left(T_{xx}(0)+T_{yy}(0)+T_{zz}(0)\right),\delta T_{xx}(x)\right]\right\rangle,
\end{eqnarray}
where the brackets describe a thermal average, $\langle A\rangle\equiv {\rm Tr} Ae^{-\beta H}$, the time dependence of the operators are defined by $A(t)\equiv e^{iHt}Ae^{-iHt}$, and the deviations from vacuum expectations are denoted by $\delta A\equiv A-\langle A\rangle$.

Coefficients can be calculated reliably in perturbation theory at high temperature when the coupling is small \cite{perturbation,Jeon:1995zm}. Unfortunately, lattice calculations, which should be valid for large coupling, do not yet provide reliable values for the viscosity \cite{sakai}. The difficulty with calculating the correlations above from lattice calculations derives from the fact that the operators are evaluated for real times, whereas the path integral used in a thermal trace samples a path over imaginary time,
\begin{eqnarray}
Z&=&\int d\vec{\phi}~\exp\left\{-i\int d^4x {\cal L}(\phi,\partial\phi)\right\},~~~~0<x_0<i\beta\\
\nonumber
\int d\vec{\phi}&\equiv&\int \prod_{j_xj_yj_zj_\beta} d\phi_{\vec{j},r}d\phi_{\vec{j},i}.
\end{eqnarray}
Herre, the lattice points are denoted by $j_x,j_y,j_z,j_\beta$. In order to represent a trace, the fields are cyclic in the time variable,
\begin{equation}
\phi(0,\vec{x})=\phi(i\beta,\vec{x}),
\end{equation}
while each of the spatial dimensions is confined to a length $L$. The expression becomes exact in the limit that the lattice points $\vec{j}$ become infinitely numerous. Although the expression above assumes scalar fields, more complicated expression for Fermi or for vector fields will not change any of following derivations.

Correlations, similar in appearance to those used to calculate the viscosities, can be expressed in terms of path integrals,
\begin{equation}
\label{eq:proppath}
{\cal P}\langle A(0)B(t)\rangle=\frac{1}{Z}\int d\vec{\phi}~A(0)B(t)\exp\left\{-i\int d^4x {\cal L}(\phi,\partial\phi)\right\},
\end{equation}
where the operators $A$ and $B$ can be expressed in terms of the fields, assuming that $t$ corresponds to a point on the path, and tht ${\cal P}$ orders $A$ and $B$ according to the ordering on the path. For instance, $A$ might refer to $\int d^3xT_{xy}$, where the stress-energy tensor can be expressed in terms of the fields,
\begin{equation}
T^{\alpha\beta}(x)=-{\mathcal L}(x)g^{\alpha\beta}
+\frac{\partial{\mathcal L}}{\partial^\alpha\phi}\partial_\beta\phi.
\end{equation}

The notational burden can be reduced by referring to correlations of an operator with itself as 
\begin{eqnarray}
G_\eta(t)&\equiv& \langle A_\eta(0)A_\eta(t)\rangle,\\
\nonumber
A_\eta&=&\frac{1}{\sqrt{V}}\int d^3r~ T_{xy}({\bf r}),\\
\nonumber
G_\zeta(t)&\equiv& \langle A_\zeta(0)A_\zeta(t)\rangle,\\
\nonumber
A_\zeta&=&\frac{1}{\sqrt{V}}\int d^3r~ \left[\frac{T_{xx}+T_{yy}+T_{zz}}{3}-P-c_s^2(T_{00}-\epsilon)\right].
\end{eqnarray}
In the last term, the subtraction of $c_s^2(T_{00}-\epsilon)$ enforces the correlator to vanish for large times. Otherwise, energy conservation would result in the fluctuation of the energy density (which is related to the specific heat) causing a correlation between the elements $T_{ii}$ at infinte relative time. With these definitions the transport coefficients become,
\begin{eqnarray}
\sigma_a^2&=&G_\eta(0),\\
\nonumber
\eta&=&\frac{-i}{2}\int_{-\infty}^{\infty}dt~t \left[G_\eta(t)-G_\eta(-t)\right],\\
\nonumber
\sigma_b^2&=&G_\zeta(0),\\
\nonumber
\zeta&=&\frac{-i}{2}\int_{-\infty}^{\infty}dt~t \left[G_\zeta(t)-G_\zeta(-t)\right].
\end{eqnarray}
Thus, $\sigma_a^2$ and $\sigma_b^2$ can be explicitly determined from lattice calculations because the operators are evaluated at $t=0$, which is located on the sampled path, $0<t<i\beta$. 

The remainder of this section is devoted to the prospects of determining $\eta$ and $\zeta$. In particular, we investigate the usefulness of calculating moments of the path integral,
\begin{equation}
C_{mn}(\beta)\equiv\int_{0}^{i\beta} dt_1dt_2~{\cal P}\langle A(t_1)A(t_2)\rangle t_1^mt_2^n.
\end{equation}
For the remainder of the section, the subscripts $a,b,\eta$ and $\zeta$ will be suppressed. Such moments can be calculated by considering a sampling of paths, each with a weight $w_j$. For each path, one can calculate $\gamma_{j,m}=\int dt~A(t)t^m$, which allows one to calculate
\begin{equation}
C_{mn}(\beta)=\frac{\sum_j w_j~\gamma_{j,m}\gamma_{j,n}}{\sum_j w_j}.
\end{equation}
The following analysis is aimed at evaluating how one might use $C_{mn}$ to determine the viscosity.

Using the fact that $\langle A(t_1)A(t_2)\rangle=G(t_2-t_1)$, the moments can be expressed as:
\begin{eqnarray}
C_{mn}(\beta)&=&\int_0^{i\beta}dt~G(t)\int_0^{i\beta-t} dt_1 \left[
t_1^m(t+t_1)^n+t_1^n(t+t_1)^m\right].
\end{eqnarray} 
The first several moments are:
\begin{eqnarray}
C_{00}(\beta)&=&\int_0^{i\beta}dt~ G(t)(i\beta-t),\\
\nonumber
C_{01}(\beta)(\beta)&=&\int_0^{i\beta}dt~ G(t)i\beta(i\beta-t),\\
\nonumber
C_{11}(\beta)&=&\int_0^{i\beta}dt~ G(t)\left[t(i\beta-t)^2+\frac{2}{3}(i\beta-t)^3\right],\\
\nonumber
C_{20}(\beta)&=&\int_0^{i\beta}dt~ G(t)\left[i\beta t(i\beta-t)+\frac{2}{3}(i\beta-t)^3\right].
\end{eqnarray}
Thus, $C_{01}$ provides no additional information. The reflection property of thermal propagators, 
\begin{equation}
G(t-i\beta/2)=G(i\beta/2-t),
\end{equation}
means that $G(t)$ has a reflection symmetry about the point $t=i\beta/2$. Thus, only even moments of $z\equiv t-i\beta/2$ are non-zero, and after substituting for $t=z+i\beta/2$ in the expressions above,
\begin{eqnarray}
C_{00}(\beta)&=&\frac{i\beta}{2}\int_0^{i\beta}dt~ G(t),\\
\nonumber
C_{11}(\beta)&=&-\frac{5\beta^2}{12}C_{00}+\frac{i\beta}{2}\int_0^{i\beta}dt~ z^2G(t),
\end{eqnarray}
and $C_{20}$ provides no additional information about $G(t)$. Thus, the moments $C_{mn}$ appear sufficient to determine all the  moments,
\begin{equation}
\label{eq:dmdef}
D_m(\beta)\equiv -i\int_0^{i\beta}dt~ z^mG(t),~~~z=t-i\beta/2,
\end{equation}
for even $m$ and $D_m(\beta)=0$ for all odd $m$. The moments $D_m(\beta)$, which are real, contain all the information represented by the thermal propagators when calculated for imaginary times $0<t<i\beta$. Our goal is to determine the degree to which $D_m(\beta)$ constrains the propagators for real time, and in particular, the integrated quantities in Eq. (\ref{eq:transtheory}) that represent the transport coefficients.

\begin{figure}
\centerline{\includegraphics[width=0.4\textwidth]{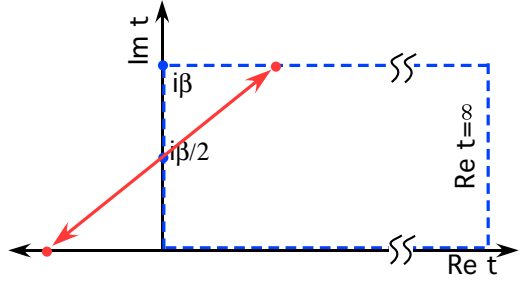}}
\caption{\label{fig:contour}(color online)
Thermal path integrals can provide $G(t)$ for $t$ along the imaginary axis. Due to analyticity, the integral of $G(t)(t-i\beta/2)^m$ along the contour (dashed line) is zero. Furthermore, the reflection properties, $G(t-i\beta/2)=G(i\beta/2-t)$ (illustrated by the arrows), permit the integral along the imaginary axis from zero to $i\beta$ to be related to the difference of the integrals along the positive and negative real axes. 
}
\end{figure}
As the transport coefficients involve integrating $G(t)$ over real time, it is insightful to express the moments $D_m$ in terms of integrals over real time. This can be accomplished by considering the dashed-line contour in Fig. \ref{fig:contour}. Whereas $D_m$ represents the integral over the left-side boundary, integrating over the entirety of the boundary should give zero. Assuming $\delta A$ is defined such that the propagator vanishes for large real times, $D_m$ can be identified with the difference in the integrals over the upper and lower boundaries. Furthermore, the integral over the upper boundary can be identified with a integral over a path with negative real time using the reflection property of $G(t)$ about the point $t=i\beta/2$. Thus,
\begin{eqnarray}
\label{eq:dm}
D_m(\beta)&=&-i\int_0^\infty dt~G(t)(t-i\beta/2)^m
+i(-1)^m\int_{-\infty}^0 dt~G(t)(t-i\beta/2)^m\\
\nonumber
&=&-i\int_0^\infty dt~\left[G(t)(t-i\beta/2)^m-G(-t)(t+i\beta/2)^m\right]\\
\nonumber
&=&2\int_0^\infty dt~\left\{G_I(t)\Re (t-i\beta/2)^m
+G_R(t)\Im (t-i\beta/2)^m\right\},\\
\label{eq:chidef}
G_I(t)&\equiv&\frac{-i}{2}[G(t)-G(-t)],~~G_R\equiv\frac{1}{2}[G(t)+G(-t)].
\end{eqnarray}
The functions $G_R$ and $G_I$ correspond to the real and imaginary parts of $G(t)$ for real $t$. It is instructive to view the last relation for the first few $m$,
\begin{eqnarray}
\label{eq:kharzeev}
D_0&=&2\int_0^\infty dt~G_I,\\
\nonumber
D_1&=&2\int_0^\infty dt~\left\{tG_I(t)-(\beta/2)G_R(t)\right\},\\
\nonumber
D_2&=&2\int_0^\infty dt~\left\{(t^2+\beta^2/4)G_I(t)-\beta tG_R(t)\right\}.
\end{eqnarray}
From inspecting the expressions above, it is clear that the moments $D_1, D_2\cdots$, are insufficient to completely determine all the moments of both $G_I$ and $G_R$. In particular, the transport coefficients, $\eta$ and $\zeta$, are defined by the integral $\int dtG_I(t)t$, which is ambiguous given that $D_1$ also depends on $G_R$. The inability to determine all the moments of interest is expected given that $G(t)$ was only determined for $0<t<i\beta$. If $G$ had been known for all $t$ along the imaginary axis, it would permit the determination of the Fourier transform of $G$, which could then be used to determine $G$ along the real axis by analytic continuation. In Reference \cite{kharzeev} the expression for $D_0$ in Eq. (\ref{eq:kharzeev}) is exploited to make a link between lattice results and viscosity, by assuming that $\int_0^\infty dt~G_I$ is related to the viscosity by a characteristic relaxation time, assumed to be the inverse temperature. However, near $T_c$ the relaxation time could be much larger, since there is little thermodynamic motivation for the system to move towards the absolute equilibrium given the broad minimum to the free energy associated with competing phases \cite{paech}.

Given the limitations discussed above, determining the transport coefficients will rely on assuming a functional form for $G(t)$, whose parameters can be fit to the moments $D_m(\beta)$. The functional form can then be applied to determine the transport coefficients. In the next three sub-sections three forms are explored. The first is based on Gaussians modified by Hermite polynomials, and the second on an exponential form. The third form \cite{sakai,karschwyld} is based on a Breit-Wigner form motivated by a single mode which oscillates with a given frequency and width.  An alternative to using moments of $G(t)$ to determine parameters for a fitting function is to use $G(t)$ at specific values of $t$ along the imaginary axis. These are also addressable with path integrals by setting $t$ to the discrete values represented by the lattice. 

\subsection{Expanding $G(t)$ with a Gaussian and Hermite polynomials}

Given the symmetry constraints, $G(t-i\beta/2)=G(-(t-i\beta/2))$, and the constraint that $G$ must vanish for large real time, Gaussians, and the associated orthogonal polynomials seem a reasonable basis to expand,
\begin{equation}
G(t)=e^{-z^2/\tau^2}
\left[g_0+g_2H_2(z/\tau)+g_4H_4(z/\tau)\cdots\right]e^{-\beta^2/4\tau^2},
~z\equiv t-i\beta/2,
\end{equation}
where $H_i$ are Hermite polynomials, with the odd polynomials being set to zero from symmetry. The first few even polynomials are:
\begin{eqnarray}
\label{eq:hermitedef}
H_0&=&1,~H_1(x)=2x,~H_2(x)=4x^2-2,~H_3(x)=8x^3-12x,~H_4(x)=16x^4-48x^2+12,\\
\nonumber
x&=&\frac{1}{2}H_1(x),~x^2=\frac{1}{4}\left(H_2(x)+2H_0(x)\right),~x^3=\frac{1}{8}\left(H_3(x)+6H_1(x)\right),~x^4=\frac{1}{16}\left(H_4(x)+12H_2(x)+12H_0(x)\right).
\end{eqnarray}
Given the polynomials, one can perform the integrals in Eq. (\ref{eq:dmdef}) to determine the moments$D_m(\beta)$,
\begin{equation}
D_m(\beta)=-i\int_{-i\beta/2}^{i\beta/2} dz~\sum_n g_n z^mH_n (z)\exp\left\{\frac{-z^2-\beta^2/4}{\tau^2}\right\}.
\end{equation}
By expanding $z^mH_m$ in powers of $z$ using the first line of Eq. (\ref{eq:hermitedef}), then rewriting each term as a sum over $H_m$ using the second line of Eq. (\ref{eq:hermitedef}), one then can perform the integrals by exploiting the relations,
\begin{eqnarray}
I_{n}(x)&\equiv&-i\int_{-ix}^{ix}dy~e^{-y^2}H_{n}(y),\\
\nonumber
I_0(x)&=&-i\pi^{1/2}{\rm erf}(ix),\\
\nonumber
I_{n>0}(x)&=&-i\left[H_{n-1}(0)-e^{x^2}H_{n-1}(ix)\right].
\end{eqnarray}
If the expansion of $G(t)$ is second order, i.e. $g_{n\ge 4}=0$,
\begin{eqnarray}
\label{eq:dexpansion}
D_0&=&\tau\left(g_0I_0+g_2I_2\right)e^{-x^2},\\
\nonumber
D_2&=&-\frac{\tau^3}{2}\left\{g_0\left(I_0+I_2/2\right)+g_2\left(4I_0+5I_2+I_4/2\right)\right\}e^{-x^2},\\
\nonumber
D_4&=&\frac{\tau^5}{2}\left\{g_0\left(3I_0/2+3I_2/2+I_4/8\right)
+g_2\left(9I_0+33I_2/2+7I_4/2+I_6/16\right)
\right\}e^{-x^2},
\end{eqnarray}
where $I_n$ are evaluated at $I_n(x\equiv\beta/2\tau)$. Thus, if lattice calculations were to extract $D_0$, $D_2$ and $D_4$, one could then find the three parameters $\tau$, $g_0$ and $g_2$. Including the $g_4$ term in the expansion of $G(t)$ would then require $D_6$ to constrain the parameters.

The simplest expansion would be to keep only the zero'th order Hermite polynomial, i.e., $g_{n\ge 2}=0$. In that case, one would only require knowledge of $D_0$ and $D_2$ to determine the two parameters $g_0$ and $\tau$. For this case, as can be seen from inspecting Eq. (\ref{eq:dexpansion}), $x\equiv \beta/2\tau$ is determined by the ratio,
\begin{equation}
\frac{D_2}{(\beta/2)^2D_0}=-\frac{2I_0(x)+I_2(x)}{x^2 I_0(x)}.
\end{equation}
This allows $x$ to be extracted by inverting the ratio. However, the r.h.s. of the equation never falls below $1/3$ if only zero$^{\rm th}$ order polynomials are used. Thus, if lattice results were to yield a lower ratio for the left-hand side, the expansion of $G(t)$ would require more Hermite polynomials. This limit comes from the fact that, when keeping on the zero$^{\rm th}$ term, $G(t)$ has a minimum at $t=i\beta/2$ when traveling along the imaginary axis. This constraint is lifted by keeping additional terms.

The transport coefficients are defined by the integrals along the real axis of $G_I$, the imaginary part of $G(t)$, as defined in Eq. (\ref{eq:chidef}). The viscosities become (using $\eta$ to denote either the shear or bulk viscosity)
\begin{eqnarray}
\label{eq:etaintegral}
\eta&=&-\frac{1}{2}\int_{-\infty}^\infty dt~t G_I(t)=\frac{i}{2}\int_{-\infty}^\infty dt~t G(t)\\
\nonumber
&=&\int_{-\infty-i\beta/2}^{\infty-i\beta/2} dz~ (z+i\beta/2)
\left[g_0 H_0(z/\tau)+g_2H_2(z/\tau)+g_4H_4(z/\tau)\cdots\right]e^{-(z^2+\beta^2/4)/\tau^2},
\end{eqnarray}
where $z\equiv t-i\beta/2$. From analyticity, the integral will not be changed by altering the limits on the path to $\pm \infty$. After recognizing $(z+i\beta/2)=(\tau H_1+i\beta H_0)/2$, one can use orthogonality properties of the Hermite polynomials then allow one to ignore all the terms except those involving $H_0^2$ and obtain
\begin{equation}
\label{eq:etagauss}
\eta=\frac{\sqrt{\pi}}{2}\beta\tau g_0e^{-\beta^2/4\tau^2}.
\end{equation} 
The fluctuations become
\begin{equation}
\label{eq:alphagauss}
\sigma^2=G(0)=g_0+g_2H_2(-i\beta/2\tau)+g_4H_4(-i\beta/2\tau)+\cdots
\end{equation}
The fluctuation $G(t=0)$ can also be calculated directly from lattice calculations, without taking moments. 

One can see from Eq.s (\ref{eq:etagauss}) and (\ref{eq:alphagauss}) that the ratio $\eta/\sigma^2$ is limited if one were to expand $G_0$ with only zero$^{\rm th}$ order Hermite polynomials. This is related to the inability of such an expansion to cover all possibilities of $D_2/D_0$ as mentioned earlier. 

As with most expansions, the validity is determined by the convergence. If the resulting transport coefficients do not appreciably change as the number of Hermite polynomials are increased, the results should be considered robust.

\subsection{Exponential expansions}

The most intuitive form for $G(t)$ would involve exponential decay for large times. Such a form was considered in \cite{kharzeev}. Choosing such a form to satisfy the reflection properties around $t=i\beta/2$,
\begin{eqnarray}
\label{eq:gexp}
G(t)&=&\frac{A\Gamma}{2\pi}\int d\omega~\frac{\cos\omega(t-i\beta/2)}{\omega^2+\Gamma^2},\\
\nonumber
&=&A\Theta(\Re t)e^{-\Gamma (t-i\beta/2)}
+A\Theta(-\Re t)e^{\Gamma (t-i\beta/2)}.
\end{eqnarray}
The result has the desired exponential behavior, but is non-analytic along the imaginary axis, precisely where the thermal path integrals are evaluated. One can forge ahead and calculate the moments $D_m$ by assuming that $G(t)=(1/2)(G(t+\epsilon)+G(t-\epsilon))$ for $\Re t=0$,
\begin{eqnarray}
D_{m=0,2,4\cdots}=A\int_{-\beta/2}^{\beta/2}dx~(ix)^m\cos(\Gamma x).
\end{eqnarray}
Due to the non-analyticity of $G(t)$ the odd moments, which are zero when integrating along the imaginary axis, are non-zero when integrating an amount $\epsilon$ on either side of the axis. However, the calculation of the even moments are unchanged by the infinitesimal translation. Thus, only the $m=0,2,4\cdots$ thermal moments can be used to constrain the corresponding moments along the real axis.

The first two even-numbered moments are:
\begin{equation}
D_0=\frac{2A}{\Gamma}\sin(\beta\Gamma/2),~~D_2=-\frac{4A}{\Gamma^3}\sin(\beta\Gamma/2).
\end{equation}
The decay width $\Gamma$ is thus obtained by taking the ratio of the first two moments,
\begin{equation}
\Gamma^2=-\frac{2D_0}{D_2}.
\end{equation}
The viscosities are straight-forward to find by integrating the form of Eq. (\ref{eq:gexp}) over real time as was done in Eq. (\ref{eq:etaintegral}),
\begin{equation}
\eta=\frac{2A}{\Gamma^2}\sin(\beta\Gamma/2).
\end{equation}

One could increase the complexity of the form for $G(t)$ in Eq. (\ref{eq:gexp}) by incorporating an expansion in exponentials with width $\Gamma,2\Gamma\cdots$, with a similar tact as was used for Hermite polynomials in the preceding section. However, it is difficult to assess the inherent error in using a form which is expressly non-analytic. It may well turn out that the exponential form is valid if the characteristic lifetime, $1/\Gamma$ is much larger than $\beta$. This corresponds to the classical limit, in that $\Gamma\beta\rightarrow 0$ as $\hbar\rightarrow 0$. Furthermore, if one calculates the effective lifetime, $\eta/\sigma^2$ for the Gaussian/Hermite expansion in Eq.s (\ref{eq:etagauss}) and (\ref{eq:alphagauss}), it appears that that expansion might have difficulty in the opposite limit. Thus, different forms might become preferable depending on whether the characteristic lifetimes are large compared to the inverse temperature.

\subsection{Breit-Wigner Form}
\label{sec:bw}

In \cite{sakai} the shear viscosity was determined from lattice calculations by investigating $G(t)$ for imaginary $t$ and fitting to a Breit-Wigner form proposed in \cite{karschwyld}. This form has a physical motivation in that $G(t)$ corresponds to a thermal Green's function for a single mode. Using simple Bose creation/destruction operators,
\begin{eqnarray}
G(t)&=&A\left\langle (a+a^\dagger)(a(t)+a^\dagger(t))\right\rangle\\
\nonumber
&=&A\left[f(m)e^{-imt}+(1+f(m))e^{imt}\right],\\
\nonumber
&=&A\left[f(m)e^{-imt}-f(-m)e^{imt}\right].
\end{eqnarray}
where $f(m)=\langle a^\dagger a\rangle=1/(e^{\beta m}+1)$ is the Bose occupation factor. This can be rewritten in terms of a spectral function $\rho(\omega)=A\delta(\omega-m)-A\delta(-\omega-m)$,
\begin{equation}
\label{eq:gkarschdef}
G(t)=\int d\omega~\frac{e^{-\beta\omega}}{1-e^{-\beta\omega}}\rho(\omega)e^{-i\omega t}.
\end{equation}
The Karsch-Wyld assumption \cite{karschwyld} for $\rho(\omega)$ is a Breit-Wigner form,
\begin{equation}
\rho(\omega)=\frac{A}{\pi}\left[\frac{\gamma}{(m-\omega)^2+\gamma^2}-\frac{\gamma}{(m+\omega)^2+\gamma^2}\right].
\end{equation}
It is straight-forward to calculate the viscosity from this form,
\begin{eqnarray}
\eta&=&\lim_{\omega\rightarrow 0} \frac{G(\omega)-G(-\omega)}{2\omega}\\
\nonumber
&=&\frac{4A\gamma m}{(\gamma^2+m^2)^2}.
\end{eqnarray}

In order to calculate the moments $D_m$, one can rearrange $G(t)$ to the form,
\begin{equation}
\label{eq:gkarsch}
G(t)=\int d\omega~\frac{\cos(\omega(t-i\beta/2))}{\sinh(\beta\omega/2)} \rho(\omega).
\end{equation}
By comparing this form to the equivalent form for the exponential form in Eq. (\ref{eq:gexp}), one can see that the two expressions differ by the substitution
\begin{equation}
\frac{1}{\omega^2+\Gamma^2}\rightarrow \frac{\rho(\omega)}{\sinh(\beta\omega/2)}.
\end{equation}
To finish finding $D_m$, one inserts the expression for $G(t)$ into the definition for $D_m$ in Eq. (\ref{eq:dmdef}) and obtain,
\begin{eqnarray}
\label{eq:kwdm}
D_0&=&\int d\omega~\frac{\rho(\omega)}{\sinh(\beta\omega/2)}\int_{-\beta/2}^{\beta/2}dz~\cosh(\omega z)
=\frac{2Am}{m^2+\gamma^2},\\
\nonumber
D_2&=&-\int d\omega~\frac{\rho(\omega)}{\sinh(\beta\omega/2)}\int_{-\beta/2}^{\beta/2}dz~z^2\cosh(\omega z)\\
\nonumber
&=&A\Re\left\{
8(m+i\gamma)^{-3}-4\beta(m+i\gamma)^{-2}\frac{\cosh(\beta(m+i\gamma)/2)}
{\sinh(\beta(m+i\gamma)/2)}
\right\}\\
\nonumber
&&+\Im\sum_{n=1,\infty}\frac{8\pi A\rho(\omega_n)}{\omega_n^2},
~~\omega_n=2n\pi i/\beta.
\end{eqnarray}
Since the Karsch-Wyld form is based on three parameters ($A$, $m$ and $\gamma$), one must either calculate $D_4$, or use $\sigma^2=G(0)$.
\begin{equation}
\label{eq:kwsigma}
\sigma^2=2A\Re\left\{\coth(\beta(m+i\gamma)/2)\right\}
-\frac{4A\pi}{\beta}\Im\sum_{n=1,\infty}\rho(\omega_n).
\end{equation}
Rather than fitting the moments, and $G(t=0)$, one could alternatively find the best fit to $G(t)$ evaluated at the lattice points, as was done in \cite{sakai}.

Like the exponential form of the previous section, the Breit-Wigner form is non-analytic. This can be seen by calculating $G_I(t)$, which unlike $G_R$, has a simple form. From Eq. (\ref{eq:gkarschdef}),  
\begin{eqnarray}
\label{eq:kwchipp}
G_I(t)&=&\frac{-i}{2}\left(G(t)-G(-t)\right)=\frac{1}{2}\int d\omega~\rho(\omega)\sin(\omega t)\\
\nonumber
&=&2Ae^{-\gamma t}\sin(mt)\Theta(\Re t)+2Ae^{\gamma t}\sin(mt)\Theta(-\Re t).
\end{eqnarray}
Although the form is not analytic for $t$ along the imaginary axis, it is continuous, unlike the exponential form of the previous section. This maintains the condition $D_{m={\rm odd}}=0$ for the path illustrated in Fig. \ref{fig:contour}. The most distinct difference that separates the Breit-Wigner from the exponential form of the previous section is in the oscillation described by the $\sin(mt)$ term in Eq. (\ref{eq:kwchipp}). An example where such an oscillation might be physical concerns the bulk viscosity arising from a non-equilibrium mean field, which might be under-damped. The trace of the stress-energy tensor might then oscillate along with the field. It is more difficult to motivate oscillatory behavior for the shear terms in the stress energy tensor (the terms $a_i$ in the first section), which in the classical limit, should decay exponentially. One way to mimic exponential behavior would be to set $m=0$, which would lead to $G_I(t)\sim te^{-\gamma t}$.

\subsection{Summary of Parameterizations}
\label{sec:parameterizationsummary}

Path integrals performed along the imaginary axis can explicitly provide $\sigma^2_a$ and $\sigma^2_b$, which when given the viscosities $\eta$ and $\zeta$, yield the IS relaxation times through Eq.(\ref{eq:alphabetaresult}). Determining $\eta$ and $\zeta$ requires fitting to a functional form, the choice of which should be motivated by the expected physical behavior. For instance, if one does not expect oscillatory behavior of the relevant elements of the stress-energy tensor, one should avoid the Breit-Wigner form. A second consideration concerns the analyticity of $G(t)$ for $t$ along the imaginary axis. Since the time dependence is driven by the evolution operator, $e^{iHt}$, it should be analytic. Furthermore, the characteristic times for any changes in $G(t)$ should not be shorter than microscopic time scales. It is difficult to assess the dangers of using explicitly non-analytic forms, such as the exponential and Breit-Wigner forms, but intuitively one would expect the errors to be important if the time scales are small. For long time scales, one would not expect non-analyticities, which are unphysical if one is considering the short-time behavior of $G_I(t)$, to be important. 

One limit in which the various parameterizations can be compared is the classical limit. In that limit \cite{forster},
\begin{equation}
e^{\beta H} A(t) e^{-\beta H}\approx A(t)+i\beta\hbar\partial_tA(t)+{\cal O}\hbar^2,
\end{equation}
which allows one to approximate $G_I$ as
\begin{equation}
-i\langle [A,A(t)]\rangle \approx \beta\partial_t\langle AA(t)\rangle +{\cal O}\hbar.
\end{equation}
With this approximation,
\begin{eqnarray}
\eta&=&-i\int_0^\infty dt~t \langle [A,A(t)]\rangle \approx \beta \int_0^\infty dt~G(t),\\
\nonumber
D_0&=&\int_0^\infty dt~\langle [A,A(t)]\rangle\approx \beta\sigma^2.
\end{eqnarray}
The classical approximation works whenever $\beta$ is sufficiently small that when multiplied by characteristic microscopic energy scales, one should get a number much less than unity. This should be true for either large temperatures or long relaxation times, i.e., $T\tau>>\hbar$. The former expression is satisfied by all three forms in Table \ref{table:parameters} as $\beta\rightarrow 0$. One way to check the validity of the classical limit is to compare $\sigma^2$ to $D_0$, both of which are calculable on the lattice.
\begin{table}
\begin{tabular}{|c|c|c|c|}\hline
& Hermite & Exponential & Breit-Wigner \\ \hline
$G(z)$& $e^{-[z^2+(\beta/2)^2]/\tau^2}\sum_{n=0,2\cdots} g_nH_n(z)$ & $Ae^{-\Gamma z}\Theta(\Re z)
+Ae^{\Gamma z}\Theta(-\Re z)$ & $A\int d\omega\rho(\omega)e^{-i\omega z}/\sinh(\beta\omega/2)$ \\
parameters & $\tau,g_0,g_2\cdots$ & $A,\Gamma$ & $A,m,\gamma$ \\ 
$D_0$ & $2\tau e^{-x^2}\sum_ng_nI_n(x),~x=\beta/2\tau$ & $2A\sin(\beta\Gamma/2)/\Gamma$ & $2Am/(m^2+\gamma^2)$ \\
$D_2$ & Eq. (\ref{eq:dexpansion}) & $-4A\sin(\beta\Gamma/2)/\Gamma^3$ & Eq. (\ref{eq:kwdm})\\
$\sigma^2$ & $\sum_n g_nH_n(-ix)$ & $A\cos(\beta\Gamma/2)$ & Eq. (\ref{eq:kwsigma})\\
$\eta$ & $(\sqrt{\pi}/2)\beta\tau g_0e^{-x^2}$ & $2A\sin(\beta\Gamma/2)/\Gamma^2$ & $4Am\gamma/(\gamma^2+m^2)^2$\\ \hline
\end{tabular}
\caption{\label{table:parameters}
Three functional forms for $G(z=t-i\beta/2)$ are given above. In principle, lattice calculations can explicitly provide the variances $\sigma^2$ for the shear- and bulk-related components of the stress-energy tensor, as well as moments of the Green's function, $D_m$ defined in Eq. (\ref{eq:dmdef}), integrated along the imaginary axis. Fitting either these moments, or the actual values of $G$ at the lattice points, to a given functional form, can then provide a value for the viscosity. However, the value could vary substantially from form to another.
}
\end{table}

In the classical limit, all three forms provide different answers for the viscosities given the same values for $D_m$. For instance, if one uses a strictly Gaussian expression for the Hermite expansion, $g_{m>0}=0$, one can find that as $\eta\rightarrow 0$, $\eta\rightarrow \sqrt{-D_0D_2\pi/24}$, whereas for the exponential form $\eta\rightarrow\sqrt{-D_0D_2/2}$. Thus given the same values $D_0$ and $D_2$ from the lattice, the two forms would lead to viscosities that differ by nearly a factor of two. The Breit-Wigner form can vary even more, since the addition of a third parameter makes $\eta$ under-constrained if one uses only $D_0$ and $D_2$. Physically, the oscillatory nature of the Breit-Wigner form makes the moments less predictable, and underscores the importance of selecting a non-oscillatory functional form if there is physical justification. Of course, the Hermite expansion can mimic oscillatory behavior of $G(t)$ if given a sufficient number of terms.

One feature of all three functional forms described above is that they can accommodate arbitrarily short relaxation times. In some of the arguments used to motivate the KSS limit, $\eta/s>\hbar/4\pi$  \cite{kss,cohen}, appeals have been made to the energy/time uncertainty principle in the form $T\tau\gtrsim \hbar$. This constraint is easily violated by all three functional forms considered here.

\section{Summary}
\label{sec:summary}

The considerations presented here all seem to solidify Israel-Stewart approaches as the preferred approach for hydrodynamic descriptions of high-energy heavy-ion collisions. Israel-Stewart treatments are more numerically stable and physical. Although the treatments require additional parameters, they are uniquely determined by the microscopic properties of the medium, which are formally stated in a way that can be  addressed in sophisticated approaches like lattice gauge theory. Israel-Stewart parameters are only uniquely determined for small deviations from equilibrium, even if the velocity gradients are large, but treatments can be phenomenologically extended to describe large deviations by adding extra parameters to confine the magnitude of the deviation of the stress-energy tensor to physical values. 

Since lattice calculations do not directly explore thermal expectations of operators evaluated at real times, neither the viscosities or relaxation times are directly accessible. However, the ratios of the viscous parameters to the relaxation times, which correspond to equal-time thermal fluctuations of the stress-energy tensor, can be explicitly determined. Extracting the viscosities requires fitting the behavior of the thermal correlators, which can be directly determined for imaginary times, $0<t<i\beta$, to a chosen functional form, which is then analytically continued to real times. The only firm lessons to be learned from the comparison of the three functional forms investigated in Sec. \ref{sec:pathintegrals} is that the choice of form can significantly affect the answer, and that the choice of form should depend on whether one expects oscillatory behavior, and whether the classical approximation (long relaxation times compared to the inverse temperature) is justified. 

\section*{Acknowledgments}
Enlightening discussions with Azwinndini Muronga are gratefully acknowledged. Support was provided by the U.S. Department of Energy, Grant No. DE-FG02-03ER41259.

\end{document}